\documentclass{article}

\usepackage{layout,amsmath,amssymb}
\begin{document}

\title{\bf Conformal Ricci and Matter Collineations for Two Perfect Fluids}

\author{\small{M. Sharif \thanks{msharif@math.pu.edu.pk} and Naghmana Tehseen}\\
\small{Department of Mathematics, University of the Punjab,}\\
\small{Quaid-e-Azam Campus, Lahore-54590, Pakistan.}}
\date{}
\maketitle

\begin{abstract}
Conformal Ricci and conformal matter collineations for the
combination of two perfect fluids in General Relativity are
investigated. We study the existence of timelike and spacelike
conformal Ricci and matter collineations by introducing the
kinematical and dynamical properties of such fluids and using the
Einstein field equations. Some recent studies on conformal
collineations are extended and new results are found. It is worth
mentioning that we recover all the previous results a special cases.
\end{abstract}

{\bf Keywords}: Conformal collineations, Two perfect fluids.\\
{\bf PACS:} 04.20.Jb Exact solutions.
%\begin{multicols}{2}

\section{Introduction}

The Einstein field equations (EFEs), whose fundamental constituent
is the spacetime metric $g_{ab}$, are highly non-linear partial
differential equations, and therefore it is very difficult to obtain
their exact solutions. Symmetries of the geometrical/physical
relevant quantities of the General Relativity (GR) theory are known
as {\it collineations}. In general, these can be represented as
$\pounds_{\xi} {\cal A} = {\cal B}$, where ${\cal A}$ and ${\cal B}$
are the geometric/physical objects, $\xi$ is the vector field
generating the symmetry, and $\pounds_{\xi}$ signifies the Lie
derivative operator along the vector field $\xi$.

A one-parameter group of conformal motions generated by a {\it
conformal Killing vector} (CKV) $\xi$ is defined as \cite{1}
\begin{equation}\label{1.1}
\pounds_{\xi} g_{ab} = 2 \psi g_{ab},
\end{equation}
where $\pounds$ is the Lie derivative operator, $\psi=\psi(x^a)$ is
a conformal factor. If $\psi_{;\, a b} \neq 0$, the CKV is said to
be {\it proper}. Otherwise, $\xi$ reduces to the special {\it
conformal Killing vector} (SCKV) if $\psi_{;\,ab} = 0$, but
$\psi_{,a}\neq 0$. Other subcases are a {\it homothetic vector} (HV)
if $\psi_{,a}=0$ and a {\it Killing vector} (KV) if $\psi= 0$.

Duggal introduced a new symmetry called a Ricci inheritance
collineation defined by \cite{2}
\begin{equation}\label{1.2}
\pounds_{\xi} R_{ab} = \alpha R_{ab},
\end{equation}
where $\alpha =\alpha(x^a)$ is a scalar function. We shall use the
term conformal Ricci collineation (CRC) onward which reduces to
Ricci collineation (RC) for $\alpha=0$. Similarly, we can define a
matter inheritance collineation or conformal matter collineation
(CMC) by
\begin{equation}\label{1.3}
\pounds_{\xi} T_{ab} =\alpha T_{ab},
\end{equation}
where $T_{ab}$ is the energy-momentum tensor, and this becomes a
matter collineation (MC) when $\alpha=0$. The function $\alpha$ is
called the inheriting or the conformal factor.

Recently, there has been a keen interest in the study of the CKVs
and affine conformal vectors (ACVs) in a class of fluid spacetimes.
Herrera {\it et al}.~\cite{3} investigated CKVs with anisotropic
fluids. Duggal and Sharma \cite{4} extended this work to the more
general case of special ACV. Mason and Maartens \cite{5} studied the
kinematics and dynamics of conformal collineations with the general
class of anisotropic fluids and no energy flux. Duggal \cite{2,6}
discussed curvature inheritance symmetry and timelike CRCs in a
perfect fluid spacetime. Yavuz and Yilmaz {\it et al}.~\cite{7,8}
considered inheriting conformal and SCKVs and worked on the
curvature inheritance symmetry in string cosmology. Yilmaz \cite{9}
also considered timelike and spacelike collineations in a string
cloud. Baysal {\it et al}.~\cite{10} worked on spacelike CRCs in
models of a string cloud and string fluid stress tensor. Mason and
Tsamparlis \cite{11} investigated spacelike CKVs in a spacelike
congruence. Sariddakis and Tsamparlis \cite{12} studied the
applications for spacelike CKVs and matter described either by a
perfect fluid or by an anisotropic fluid. Tsamparlis \cite{13} has
discussed the general symmetries of string fluid spacetime. Sharif
and Umber \cite{14} have investigated timelike and spacelike CMCs
for specific forms of the energy-momentum tensor. In a very recent
paper, Sharif and Naghmana \cite{15} have explored the spacetimes
admitting CRCs and CMCs in anisotropic fluids.

In this paper, we consider timelike and spacelike vectors with CRCs
and CMCs for the two perfect fluids. We shall discuss the conditions
imposed in each case. The layout of the paper is the following. In
Section $2$ we review the $1+1+2$ decomposition and consider the
decomposition of the quantities which will be used in later
sections. Section $3$ contains the study of the energy-momentum
tensor for the two perfect fluids. We study collineations of the
form $\xi^a=\xi u^a,~\xi^a=\xi v^a,~\xi^a=\xi x^a$ and $\xi^a=\xi
y^a$, where $u^a,~v^a$ are the four velocities of the two perfect
fluids and $x^a,~y^a$ are unit vectors in the direction of the two
perfect fluids. We do not assume that the cosmological constant
vanishes. In Section $4$, we investigate the kinematic conditions
for timelike and spacelike CRCs and CMCs. Section $5$ is devoted to
deriving the necessary and sufficient conditions for the two perfect
fluid spacetimes which admits CRCs. In Section $6$, we derive the
necessary and sufficient conditions for the two perfect fluid
spacetimes admitting CMCs. Finally, Section $7$ contains a summary
and discussion of the results.

\section{Notations and Some General Results}

We consider the timelike vectors $u^a,~v^a$ and spacelike vectors
$x^a,~y^a$ corresponding to the combination of two perfect fluids
satisfying the following relations:
\begin{eqnarray}\label{2.4}
&&u^au_a=-1=v^av_a,\quad x^ax_a=1=y^ay_a,\nonumber\\
&&u^av_a\neq0=x^ay_a,\quad u^ax_a=0=v^ax_a,\nonumber\\
&&u^ay_a=0=v^ay_a.
\end{eqnarray}
The projection tensors corresponding to the vectors $u^a$ and $v^a$
are defined as follows:
\begin{equation}\label{2.5}
h_{1ab}=g_{ab}+u_au_b, \quad h_{2ab}=g_{ab}+v_av_b
\end{equation}
which project normal to $u^a,~v^a$ and produce the well-known $1+3$
decomposition of the tensor algebra along $u^a,~v^a$ respectively.
The two well-known examples of the $1+3$ decomposition \cite{16} are
\begin{eqnarray}\label{2.6}
&&u_{a;b}=\sigma_{1ab}+\omega_{1ab}+\frac{1}{3} \theta
h_{1ab}-\dot{u}_au_b,\\\label{2.7}
&&v_{a;b}=\sigma_{2ab}+\omega_{2ab}+\frac{1}{3} \theta
h_{2ab}-\dot{v}_av_b,\\\label{2.8}
&&T_{ab}=\mu_1u_au_b+\mu_2v_av_b+F_1 h_{1ab}+F_2
h_{2ab}\nonumber\\&&+2q_{1(a}u_{b)}+\pi_{1ab}+2q_{2(a}v_{b)}+\pi_{2ab}.
\end{eqnarray}
In this paper we shall use the following notation:\\
For the fluid 1, when $\dot{u}_a=u_{a;b}u^b$, the notation $u_{a;b}$
means derivative w.r.t.~$x$ but for $\hat{u}_a=u_{a;b}u^b$,
$u_{a;b}$ denotes derivative w.r.t.~$y$. Similarly, for the fluid 2,
when we use $\dot{v}_a=v_{a;b}v^b$, the symbol $v_{a;b}$ indicates
derivative w.r.t.~$y$ and for $\hat{v}_a=v_{a;b}v^b$, $v_{a;b}$
means derivative w.r.t.~$x$.

The two pair of vectors $u^a,~x^a$ and $v^a,~y^a$ define projection
operators as
\begin{equation}\label{2.9}
H_{1ab}=h_{1ab}-x_ax_b,\quad H_{2ab}=h_{2ab}-y_ay_b
\end{equation}
which are normal to both $u^a,~x^a$ and $v^a,~y^a$. These projection
operators satisfy the following properties
\begin{eqnarray}\label{2.10}
&&H_{1ab}u^a=0=H_{1ab}x^a,\quad H_{1ab}h_{1c}^b=H_{1ac},
\nonumber\\
&&H^a_{1a}=2=H^a_{2a}, \quad H_{2ab}v^a=0=H_{2ab}y^a,\nonumber\\
&&H_{2ab}h_{2c}^b=H_{2ac}.
\end{eqnarray}
The vectors $x_{a;b}$ and $y_{a;b}$ can be decomposed using the same
procedure as given in \cite{17}. This gives the following
\begin{eqnarray}\label{2.11}
&&x_{a;b}=A_{1ab}+x{^*_a}x_b-\dot{x}_au_b\nonumber\\
&&+u_a[x{^f{u_{f;b}}}+(x^f\dot{u}_f)u_b-(x^fu{^*_f})x_b],\\
\label{2.12}
&&y_{a;b}=A_{2ab}+y{^*_a}y_b-\dot{y}_av_b\nonumber\\
&&+v_a [y{^f{v_{f;b}}}+(y^f\dot{v}_f)v_b-(y^fv{^*_f})y_b],
\end{eqnarray}
where
\begin{eqnarray}\label{2.13}
A_{1ab}=S_{1ab}+R_{1ab}+\frac{1}{2}\varepsilon_1H_{1ab},\nonumber\\
A_{2ab}=S_{2ab}+R_{2ab}+\frac{1}{2}\varepsilon_2H_{2ab},
\end{eqnarray}
and $s^*= s_{...;a}x^a$. The notations $x_{;b}$ and $y_{;b}$ in
Eqs.~(\ref{2.11}) and (\ref{2.12}) mean derivative w.r.t~$x$ and $y$
respectively. We note that $S_{1ab}=S_{1ba}$ and $S^b_b=0$ is the
traceless part (shear tensor). Also, $R_{1ab}=-R_{1ba}$ is
antisymmetric part (rotation tensor) and $\varepsilon_1$,
$\varepsilon_2$ are the traces (expansion). The following relations
can be found \cite{17}
\begin{eqnarray}\label{2.14}
&&S_{1ab}= H^c_{1a}H^d_{1b} x_{(c;d)}-\frac{1}{2}\varepsilon_1H_{1ab},\\
\label{2.15}&&R_{1ab}=H^c_{1a} H^d_{1b} x_{[c;d]} ,\quad
\varepsilon_1=H_1^{ab}x_{a;b},\\
\label{2.16}&&S_{2ab}= H^c_{2a} H^d_{2b} y_{(c;d)}-\frac{1}{2}\varepsilon_2H_{2ab},\\
\label{2.17}&&R_{2ab}= H^c_{2a} H^d_{2b} y_{[c;d]} ,\quad
\varepsilon_2=H_2^{ab}y_{a;b}.
\end{eqnarray}
The bracket terms of $u^a$ and $v^a$ in Eqs.~(\ref{2.11}) and
(\ref{2.12}) can be written as
\begin{eqnarray}\label{2.18}
&&-N_{1b}+2w_{1fb}x^f+H^f_{1b}\dot{x_f},\nonumber\\
&&-N_{2b}+2w_{2fb}x^f+H^f_{2b}\dot{y_f},
\end{eqnarray}
where $N_{1b},~N_{2b}$ are given by
\begin{equation}\label{2.19}
N_{1b}=H^a_{1b}(\dot{x_a}-u{_a}^*),\quad
N_{2b}=H^a_{2b}(\dot{y}_a-v{_a}^*).
\end{equation}
They are called Greenberg vectors \cite{18}. This vector vanishes if
and only if the vector fields $u^a,~v^a,~x^a,~y^a$ are surface
forming, i.e., if and only if $\pounds_\xi x^a=Au^a+Bx^a$. From the
kinematic point of view, the vector $N_{1a}$ vanishes if and only if
the vector field $x^a$ is {\it frozen} along the observer.

Substituting Eq.~(\ref{2.18}) in Eqs.~(\ref{2.11}) and (\ref{2.12}),
it follows that
\begin{eqnarray}\label{2.20}
&&x_{a;b}=A_{1ab}+x^*{_a}x_b-\dot{x}_au_b+H_{1b}^c\dot{x_c}u_a+(2w_{1tc}x^t-N_{1c})u_a,\\
\label{2.21}&&y_{a;b}=A_{2ab}+y^*_ay_b-\dot{y}_au_b+H_{2b}^c\dot{y}_cv_a+(2w_{2tc}y^t-N_{2c})v_a.
\end{eqnarray}
Notice that $\pounds_\xi$ means Lie derivative with respect to the
vector field $\xi$, otherwise $\xi$ is used as a scalar.

\section{The Two Perfect Fluids}

The energy-momentum tensor for a non-interacting combination of two
(non-zero) perfect fluids is \cite{19}
\begin{equation}\label{3.1}
T_{ab}=(\rho_1+p_1)u_au_b+(\rho_2+p_2)v_av_b+(p_1+p_2)g_{ab},
\end{equation}
where $p_1,~p_2$ are pressures, $\rho_1,~\rho_2$ are densities,
$u^a,~v^a$ are the four-velocities $(u^au_a=-1=v^av_a,~u^av_a\neq
0)$ and $x^a,~y^a$ are unit spacelike vectors normal to the four
velocities $(u^ax_a=v^ax_a=0=u^ay_a=v^ay_a)$. The energy-momentum
tensor of the two fluids can also be written as
\begin{equation}\label{3.2}
T_{ab}=\rho_1u_au_b+\rho_2 v_av_b+p_1h_{1ab}+p_2h_{2ab}.
\end{equation}
Comparing Eqs.~(\ref{2.8}) and (\ref{3.2}), we obtain
\begin{eqnarray}\label{3.3}
\mu_1=\rho_1,\quad \mu_2=\rho_2, \quad F_1=p_1, \quad
F_2=p_2,\nonumber\\ \quad q_1^a=0, \quad q_2^a=0,\quad \pi_{1ab}=0,
\quad \pi_{2ab}=0.
\end{eqnarray}
This implies that the two perfect fluids is an anisotropic fluid for
which heat flux and traceless anisotropic stress tensor are zero.
The EFEs can be written as
\begin{equation}\label{3.4}
R_{ab}=T_{ab}+(\Lambda-\frac{1}{2}T)g_{ab},
\end{equation}
where $\Lambda$ is the cosmological constant. For the two perfect
fluids, it takes the form
\begin{eqnarray}\label{3.5}
&&R_{ab}=(\rho_1+p_1)u_au_b+(\rho_2+p_2)v_av_b+
\frac{1}{2}(\rho_1+\rho_2-p_1-p_2+2\Lambda)g_{ab}.
\end{eqnarray}
The $1+3$ decomposition of $R_{ab}$ takes the form
\begin{eqnarray}\label{3.6}
&&R_{ab}=\frac{1}{2}(\rho_1+3p_1-\Lambda)u_au_b+\frac{1}{2}(\rho_2+3p_2-\Lambda)v_av_b
+\frac{1}{2}(\rho_1-p_1+\Lambda)h_{1ab}\nonumber\\
&&+\frac{1}{2}(\rho_2-p_2+\Lambda)h_{2ab}.
\end{eqnarray}
This gives the field equations in terms of the combination of two
perfect fluid variables. Using Eq.~(\ref{3.6}), we find $\pounds_\xi
R_{ab}$ in terms of the two perfect fluids.
\begin{eqnarray}\label{3.7}
&&\frac{1}{\xi}\pounds
R_{ab}=[\frac{1}{2}(\rho_1-\rho_2+3p_1+p_2\dot{)} +(\rho_1-\rho_2+3p_1+p_2-2\Lambda)(\ln\xi\dot{)}]u_au_b+[(\rho_1\nonumber\\
&&-\rho_2+3p_1+p_2-2\Lambda)(\dot{u}_c-(\ln\xi)_{,
c})]u_{(a}h^c_{1b)}+\{\frac{1}{2}(\rho_1+\rho_2-p_1-p_2\dot{)}h_{1cd}+
(\rho_1\nonumber\\
&&+\rho_2-p_1-p_2+2\Lambda)(\sigma_{1cd}+\frac{1}{3}\theta_1h_{1cd}))\}h^c_{1d}h^d_{1b}
+(\rho_2+p_2\dot{)}v_av_b+2(\rho_2+p_2)
(v_fv_{(a}u^f_{;b)}\nonumber\\
&&+\dot{v}_{(a}v_{b)}+u^cv_cv_{(a}(\ln\xi)_{,b)}).
\end{eqnarray}
Similarly, the Lie derivative of the Ricci tensor along the
spacelike vector $\xi^a=\xi x^a$ can be written in terms of the
$1+1+2$ dynamic quantities.
\begin{eqnarray}\label{3.8}
&&\frac{1}{\xi}\pounds
R_{ab}=[\frac{1}{2}(\rho_1-\rho_2+3p_1+p_2{)}^*
+(\rho_1-\rho_2+3p_1+p_2-2\Lambda)\dot{u}^fx_f]u_au_b
+2[\frac{1}{2}(\rho_1+\rho_2\nonumber\\
&&-p_1-p_2+2\Lambda)u^*_fx^f
-\frac{1}{2}(\rho_1+\rho_2-p_1-p_2+2\Lambda)(\ln
\xi\dot{)}]u_{(a}x_{b)}-2[\frac{1}{2}(\rho_1+\rho_2-p_1-p_2\nonumber\\
&&+2\Lambda)N_{1d}
+(\rho_1-\rho_2+3p_1+p_2-2\Lambda)\omega_{1cd}x^c]u_{(a}H^d_{1b)}
+[\frac{1}{2}(\rho_1+\rho_2+p_1-p_2{)}^*+(\rho_1+\rho_2\nonumber\\
&&-p_1-p_2-2\Lambda)\times(\ln \xi{)}^*]
x_ax_b+2[\frac{1}{2}(\rho_1+\rho_2-p_1-p_2)\times H_{1cd}{x^*}^c+\frac{1}{2}(\rho_1+\rho_2-p_1-p_2\nonumber\\
&&-2\Lambda)\times H^f_{1d}(\ln\xi)_{,f}]x_{(a}H^d_{1b)}
+[\frac{1}{2}[(\rho_1+\rho_2-p_1-p_2{)}^*H_{1cd}+(\rho_1+\rho_2-p_1-p_2
+2\Lambda)\nonumber\\
&&\times(S_{1cd}+\frac{1}{2}\varepsilon_1H_{1cd})]H^c_{1a}H^d_{1b}
+(\rho_2+p_2{)}^*\times v_av_b+
2(\rho_2+p_2)[v^*_{(a}v_{b)}+v_fv_{(a}x^f_{;b)}].
\end{eqnarray}
Expressions (\ref{3.7}) and (\ref{3.8}) are general and hold for all
collineations and any two perfect fluids. Similarly, we can write
expressions for the second timelike and spacelike vectors.

The conservation equations for the two perfect fluids are (there
arise the following two cases)
\par\noindent
\par\noindent
{\bf Case (i):}
\begin{eqnarray}\label{3.9}
&&\dot{\rho}_1+(\rho_1+p_1)\theta_1
-(\rho_2+p_2)u^a\hat{v}_a-\dot{p}_2-(\rho_2+p_2\hat{)}u_av^a-(\rho_2+p_2)u_av^a\tau_1=0,\\
\label{3.10}&&(\rho_1+p_1)\dot{u}^c+(\rho_2+p_2\hat{)}(v^c+u^av_au_c)+(\rho_2+p_2)(\hat{v}^c+\hat{v}^fu_fu^c+u_av^a\tau_1u_c)\nonumber\\
&&+(\rho_2+p_2)\tau_2v^c+h_1^{cf}(p_1+p_2)_{;f}=0.
\end{eqnarray}
Project the second equation along $x^c$ and with $H_{1c}^a$, we get
the two equations
\begin{eqnarray}\label{3.11}
&&p^*_1+p^*_2+(\rho_1+p_1)x_f\dot{u}^f +(\rho_2+p_2)\times
x_f\hat{v}^f=0,\\\label{3.12} &&H_{1c}^a[(\rho_1+p_1)\dot{u}_c
+((\rho_2+p_2\hat{)}v_c
+(\rho_2+p_2)\tau_2)v_c+(\rho_2+p_2)\hat{v}_c+h_{1c}^a(p_1+p_2)_{;a}]=0.
\end{eqnarray}
{\bf Case (ii):}
\begin{eqnarray}\label{3.13}
&&\dot{\rho}_2+(\rho_2+p_2)\theta_2
-(\rho_1+p_1)v^a\hat{u}_a-(\rho_1+p_1\hat{)}u_av^a-(\rho_1+p_1)u_av^a\tau_2-\dot{p}_1=0,\\\label{3.14}
&&(\rho_2+p_2)\dot{v}^c+(\rho_1+p_1\hat{)}(u^c+u^av_av_c)+(\rho_1+p_1)(\hat{u}^c+\hat{u}^fv_fv^c+u_av^a\tau_2v_c)
\nonumber\\
&&+(\rho_1+p_1)\tau_1u^c+h_2^{cf}(p_1+p_2)_{;f}=0.
\end{eqnarray}
Project the second equation along $y^c$ and with $H_{2c}^a$, the
equations are
\begin{eqnarray}\label{3.15}
&&p^*_1+p^*_2+(\rho_2+p_2)y_f\dot{v}^f +(\rho_1+p_1)y_f\hat{v}^f=0,\\
\label{3.16} &&H_{1c}^a[(\rho_2+p_2)\dot{v}_c
+((\rho_1+p_1\hat{)}u_c+(\rho_1+p_1)\hat{u}_c
+(\rho_1+p_1)\tau_1)u_c+(\rho_1+p_1)\hat{u}_c+h_{2c}^a(p_1+p_2)_{;a}]=0.
\end{eqnarray}

\section{Kinematic Conditions for the Two Perfect Fluids}

Kinematics and dynamics in GR are clearly defined by considering the
kinematical and dynamical variables, and the identities and the
constraints they have to satisfy. Symmetries are an important form
of constraints which restrict a physical system. These restrictions
are expressed as relations among the parameters specifying the {\it
state} of the system. Collineations restrict the system by the two
levels, i.e., at the kinematical level (relations among the
kinematic and the geometric variables only) and at the dynamical
level (relations among the kinematic and the dynamical variables).
Here we use kinematic restrictions coming from a general
collineation, in particular, from a CRC. Since collineations are
defined in terms of the Lie derivative of the metric tensor $g_{ab}$
and its derivatives, all types of collineations can be expressed by
the quantity $\pounds_\xi g_{ab}$. We define the decomposition as
\cite{13}
\begin{equation}\label{4.1}
\pounds_\xi g_{ab}=2\psi g_{ab}+2P_{ab},
\end{equation}
where $\psi(x^a)$ is a function (the conformal factor) and
$P_{ab}(x^a)$ is a symmetric traceless tensor. This implies that
every collineation can be expressed in terms of
$\psi(x^a),~P_{ab}(x^a)$ and their derivatives. For $P_{ab}=0$, this
gives a CKV; $\psi_{;a}=0=P_{ab;c}$ yields an affine collineation,
etc. The kinematic conditions of a general collineation are
relations among the kinematic quantities (shear, rotation,
expansion) of the vector field involved and the parameters
$\psi(x^a),~P_{ab}(x^a)$.

\subsection{Timelike Collineation}

There arise the following two cases:\\
\par\noindent
\par\noindent
(i)\quad ${\xi}^a={\xi}u^a$,\quad (ii)\quad
${\xi}^a={\xi}v^a,~({\xi}\neq0)$.\\
\par\noindent
\par\noindent
{\bf Case (i):} For this case, Eq.~(\ref{4.1}) can be written in the
form
\begin{equation}\label{4.2}
\xi_{a;b}+\xi_{b;a}=2{\psi}g_{ab}+2P_{ab},
\end{equation}
where
\begin{eqnarray}\label{4.3}
&&\psi=\frac{\xi}{4}[(\ln\xi\dot{)}+\theta_1],\\\label{4.4}
&&P_{ab}= \xi[\sigma_{1ab}+ \frac{1}{3} \theta_1
h_{1ab}-\hat{u}_{(a}u_{b)}+(\ln\xi)_{,(a} u_{b)}-\frac{1}{4}
\{(\ln\xi\dot{)}+\theta_1\} g_{ab}].
\end{eqnarray}
We $1+3$ decompose $P_{ab}$ w.r.t.\ the timelike vector $u^a$ in the
following
\begin{eqnarray}\label{4.5}
&&P_{ab}=\frac{\xi}{4}[\theta_1-3(\ln\xi\dot{)}]u_au_b-
\xi[\dot{u_c}-h_{1c}^d(\ln\xi)_{,d}]\times
h_{1(a}^cu_b)+\frac{1}{12}\xi[\theta_1-3(\ln\xi\dot{)}]h_{1ab}+\xi\sigma_{1ab}.
\end{eqnarray}
If we take
\begin{eqnarray}\label{4.6}
&&\mu_{1P}=\frac{\xi}{4}[\theta_1-3(\ln\xi\dot{)}],\\
\label{4.7}&&L_{1P}=\frac{1}{12}\xi[\theta_1-3(\ln\xi\dot{)}],\\
\label{4.8}&&\Upsilon_{1P}=\xi[\dot{u}_c-h_{1c}^d(\ln\xi)_{,d}],\\
\label{4.9}&&M_{P_{1ab}}=\sigma_{1ab}
\end{eqnarray}
then Eq.~(\ref{4.5}) can take the following form
\begin{equation}\label{4.10}
P_{ab}=\mu_{1P}u_au_b+L_{1P}h_{1ab}- 2\Upsilon_{1P_{(a}}u_{b)}+\xi
M_{1P_{ab}}.
\end{equation}
We express these equations as conditions among the kinematic
variables of the timelike congruence $u^a$, which gives a system of
equations called the kinematic conditions of the collineations.\\
{\bf Case (ii):} In this case, we $1+3$ decompose the traceless
tensor $P_{ab}$ w.r.t.\ the timelike vector $v^a$. Using the same
procedure as above, it follows that
\begin{equation}\label{4.11}
P_{ab}=\mu_{2P}v_a v_b+L_{2P}h_{2ab} -2\Upsilon_{2P_{(a}}v_{b)}+\xi
M_{2P_{ab}},
\end{equation}
where
\begin{eqnarray}\label{4.12}
&&\mu_{2P}=\frac{\xi}{4} [\theta_2-3(\ln\xi\dot{)}],\\\label{4.13}
&&L_{2P}=\frac{1}{12}\xi [\theta_2-3(\ln\xi\dot{)}],\\\label{4.14}
&&\Upsilon_{2P}=\xi[\dot{v}_c- h_{2c}^d(\ln\xi)_{,d}],\\\label{4.15}
&&M_{P_{1ab}}=\sigma_{2ab}.
\end{eqnarray}
The kinematic conditions for a CRC involve the second derivatives of
the quantities $\psi,~P_{ab}$. We evaluate these conditions by
taking a general vector field $\xi^a$ with the identities
\begin{eqnarray}\label{4.16}
&&\pounds_\xi
R_{ab}=(\pounds_\xi\Gamma^c_{ab})_{;c}-(\pounds_\xi\Gamma^c_{ac})_{;b},\\
\label{4.17}
&&\pounds_\xi\Gamma^a_{bc}=\frac{1}{2}g^{ad}\{(\pounds_\xi
g_{bd})_{;c}+(\pounds_\xi g_{cd})_{;b}-(\pounds_\xi
g_{bc})_{;d}\}.
\end{eqnarray}
Using Eqs.~(\ref{4.1}), (\ref{4.16}), (\ref{4.17}), the following
general result for any vector $\xi^a$ can be verified.\\
\par\noindent
\par\noindent
\textbf{Proposition 1:} A fluid spacetime $u^a$ admits a CRC $\xi^a$
if and only if
\begin{eqnarray}\label{4.18}
\triangle\psi=\frac{1}{3}(P^{ab}_{;ab}-aR),\\
\label{4.19} \triangle P_{ab}=2K_{ab}-2A_{ab}-2aZ_{ab},
\end{eqnarray}
where $R$ is the Ricci scalar and
\begin{eqnarray}\label{4.20}
&&K_{ab}=P^c_{(a;b)c}-\frac{1}{4}g_{ab}P^{ab}_{;ab},\quad
A_{ab}=\psi_{;ab}-\frac{1}{4}g_{ab}\Delta\psi,\nonumber\\
&&Z_{ab}=R_{ab}-\frac{1}{4}g_{ab}R,
\end{eqnarray}
which is a geometric result. The kinematic conditions can be
obtained by replacing $\psi,~P_{ab}$ from Eqs.~(\ref{4.3}),
(\ref{4.4}) in terms of the kinematic variables. The resulting
expressions will become very tedious and hence will not be given
here. These are the constraints satisfied by any solution.

\subsection{Spacelike Collineation}

The kinematic restrictions in this case involve all the nine
quantities, i.e.,
\begin{eqnarray*}
\sigma_{1ab},~\omega_{1ab},~
\theta_1,~\dot{u}_a,~S_{1ab},~R_{1ab},~\varepsilon_1,~\dot{x}_a,~u^*_a,\\
\sigma_{2ab},~\omega_{2ab},~\theta_2,~\dot{v}_a,~S_{2ab},~R_{2ab},
~\varepsilon_2,~\dot{y}_a,~v^*_a,
\end{eqnarray*}
plus the parameters $\psi,~P_{ab}$ and their derivatives. Again we
have two cases according to
\par\noindent
(i)\quad ${\xi}^a={\xi}x^a$,\quad(ii)\quad${\xi}^a={\xi}y^a$.
\par\noindent
{\bf Case (i):} We make the $1+1+2$ decomposition of $P_{ab}$ by
considering Eq.~(\ref{4.1}) and contract with
\begin{eqnarray*}
&&u^a u^b,~u^a x^b,~x^a x^b,~H_{1c}^b u^a,~H_{1c}^b x^a,~H_{1ab},\\
&&H_{1c}^a H_{1d}^b-\frac{1}{2}H_1^{ab}H_{1cd},
\end{eqnarray*}
it turns out that
\begin{eqnarray}\label{4.21}
\psi=\frac{\xi}{4}[\varepsilon_1
+{(\ln\xi)}^*-\dot{x^c}u_c],\\\label{4.22}
\lambda_{1P}=P_{ab}u^au^b=\frac{\xi}{4}
[\varepsilon_1+{(\ln\xi)}^*+3\dot{x^c}u_c],\\\label{4.23}
2K_{1p}=-2P_{ab}u^ax^b
=-\xi[(\ln\xi\dot{)}+x^{*c}u_c],\\\label{4.24}
\gamma_{1P}=P_{ab}x^ax^b=-\frac{\xi}{4}
[\varepsilon_1-3{(\ln\xi)}^*-\dot{x^c}u_c],\\\label{4.25}
2S_{1Pc}=-2P_{ab}H_{1c}^bu^a
=-\xi[N_{1c}-2w_{1tc}x^t],\\\label{4.26}
2\varrho_{1Pc}=2P_{ab}H_{1c}^bu^a=\xi H_{1c}^b
[(\ln\xi)_{,b}+{x_b}^*],\\\label{4.27}
a_{1P}=P_{ab}H_1^{ab}=\frac{\xi}{2}
[\varepsilon_1-{(\ln\xi)}^*+\dot{x^c}u_c],\\\label{4.28}
D_{1P_{ab}}=\xi S_{1ab}=
(H_{1c}^aH_{1d}^b-\frac{1}{2}H_1^{ab}H_{1cd})P_{ab}.
\end{eqnarray}
Thus the $1+1+2$ decomposition of $P_{ab}$ is given by
\begin{eqnarray}\label{4.29}
P_{ab}=\lambda_{1P}u_au_b+2K_{1P}u_{(a}x_{b)} +2S_{1P(a}u_{b)}
+\gamma_{1P}
x_ax_b+2\varrho_{1P_{(a}}x_{b)}+\frac{1}{2}a_{1P}H_{1ab}+D_{1P_{ab}}.
\end{eqnarray}
The property $P^a_a=0$ implies that
\begin{equation}\label{4.30}
a_{1P}-\lambda_{1P}+\gamma_{1P}=0.
\end{equation}
\par\noindent
{\bf Case (ii):} The $1+1+2$ decomposition in this case can be
obtained by contracting Eq.~(\ref{4.1}) with
\begin{eqnarray*}
&&v^av^b,~v^ay^b,~y^ay^b,~H_{2c}^bv^a,~H_{2c}^by^a,\\
&&H^a_{2c}H_{2d}^b-\frac{1}{2}H_2^{ab}H_{2cd}.
\end{eqnarray*}
It follows that
\begin{eqnarray}\label{4.31}
\psi=\frac{\xi}{4}[\varepsilon_2+
{(\ln\xi)}^*-\dot{y^c}v_c],\\\label{4.32}
\lambda_{2P}=P_{ab}v^av^b=\frac{\xi}{4}
[\varepsilon_2+{(\ln\xi)}^*+3\dot{y^c}v_c],\\\label{4.33}
2K_{2p}=2P_{ab}v^ay^b= \xi[(\ln\xi\dot{)}+{y^c}^*v_c],\\\label{4.34}
\gamma_{2P}=P_{ab}y^ay^b=-\frac{\xi}{4}
[\varepsilon_2-3(\ln\xi{)}^*-\dot{y^c}v_c],\\\label{4.35}
2S_{2pc}=-2P_{ab}H_{2c}^bv^a=
-\xi[N_{2c}-2w_{2tc}y^t],\\\label{4.36}
2\varrho_{2Pc}=2P_{ab}H_{2c}^bv^a=\xi
H_{2c}^b[(\ln\xi)_{,b}+{y_b}^*],\\\label{4.37}
a_{2p}=P_{ab}H_2^{ab}=\frac{\xi}{2}
[\varepsilon_2-(\ln\xi{)}^*+\dot{y^c}v_c],\\\label{4.38}
D_{2P_{ab}}=\xi S_{2ab}=(H_{2c}^aH_{2d}^b
-\frac{1}{2}H_2^{ab}H_{2cd})P_{ab}.
\end{eqnarray}
The $1+1+2$ decomposition of $P_{ab}$ turns out to be
\begin{eqnarray}\label{4.39}
&&P_{ab}=\lambda_{2P}v_av_b+2K_{2P}v_{(a}y_{b)}
+2S_{2p(a}v_{b)}+\gamma_{2P}
y_ay_b+2\varrho_{P_{2(a}}y_{b)}+\frac{1}{2}a_{2P}H_{ab}+D_{2P_{ab}}.
\end{eqnarray}
From $P^a_a=0$, we have
\begin{equation}\label{4.40}
a_{2P}-\lambda_{2P}+\gamma_{2P}=0.
\end{equation}

\section{Conformal Ricci Collineations for the Two Perfect Fluids}

In this section we shall discuss the existence of timelike and
spacelike CRCs for the two perfect fluids.

\subsection{Timelike Conformal Ricci Collineations}

We shall give the necessary and sufficient conditions for the
timelike CRCs of the two perfect fluids for the following two
cases:\\
(i)\quad ${\xi}^a={\xi}u^a$,\quad (ii)\quad ${\xi}^a={\xi}v^a$.\\
From Eq.~(\ref{3.5}), we have the $1+3$ decomposition of the Ricci
tensor and hence CRC gives the condition
\begin{eqnarray}\label{5.1}
&&\pounds_\xi R_{ab}=\alpha[\frac{1}{2}(\rho_1+3p_1-\Lambda)u_au_b
+\frac{1}{2}(\rho_2+3p_2-\Lambda)v_av_b+\frac{1}{2}(\rho_1-p_1+\Lambda)h_{1ab}
\nonumber\\&&+\frac{1}{2}(\rho_2-p_2+\Lambda)h_{2ab}],
\end{eqnarray}
and the corresponding $1+3$ expression of $\pounds_\xi R_{ab}$ is
given in Eq.~(\ref{3.7}). Equating these two expressions and after
some calculation we find the following results.\\
\par\noindent
\par\noindent
\textbf{Proposition 2:} A two perfect fluid spacetime admits a CRC
${\xi}^a={\xi}u^a$ if and only if
\begin{eqnarray}\label{5.2}
&&\dot{\rho}_1=\dot{p}_2+(\rho_2+p_2)(\hat{v}^a u_a+u_av^a\tau_2)
-(\rho_1+p_1)\theta_1+(\hat{\rho}_2+\hat{p}_2)u_av^a,\\
&&\dot{p}_1=\frac{1}{3}\dot{\rho}_2-\frac{2}{3}\dot{p}_2
+(\rho_2+p_2)(\hat{v}^au_a+v_au^a\tau_2)
+(\hat{\rho}_2+\hat{p}_2)u^av_a+\frac{1}{3}(2\rho_2-\rho_1-5p_1\nonumber\\
&& -2p_2+4\Lambda)\theta_1
+\frac{2}{3}(\rho_2+p_2\dot{)}u^av_au^bv_b+\frac{4}{3}(\rho_2+p_2)\label{5.3}
\times(\dot{v}_au^av_bu^b+u^av_a(\ln\xi\hat{)}\nonumber\\
&&+u^av_au^bv_b(\ln\xi\dot{)}+v_a\hat{u}^a+v_a\dot{u}^av_bu^b)
-\frac{\beta}{3}(3\rho_1+\rho_2-3p_1-5p_2+6\Lambda+2(\rho_2+p_2)v_au^av_bu^b),\\\label{5.4}
&&(\rho_1-\rho_2+3p_1+p_2-2\Lambda)[\dot{u}_a-(\ln\xi)_{,a}
-(\ln\xi\dot{)}u_a]+2(\rho_2+p_2\dot{)}(u^cv_cv_a+u^cv_cu^dv_du_a)\nonumber\\
&&+2(\rho_2+p_2)(\dot{v}_cu^cv_a+2\dot{v}_cu^cv_du^du_a
+v_cu^c\dot{v}_a+u_cv^cu^dv_d(\ln\xi)_{,a}
+2u_cv^cu^dv_d(\ln\xi\dot{)}u_a\nonumber\\
&&+u^cv_c(\ln\xi\dot{)}v_a+v_cu^cv_du^d_{;a}
+2v_c\dot{u}^cv_du^du_a+v_c\dot{u}^cv_a)=\beta(\rho_2+p_2)(v_cu^cv_a+v_cu^cv_du^du_a),\\\label{5.5}
&&(\rho_2-\rho_1-3p_1-p_2+2\Lambda)[\theta_1-(\ln\xi\dot{)}]
+2(\hat{\rho}_2+\hat{p}_2)u^cv_c+2(\rho_2+p_2)(\hat{v}^au_a
+u_av^a\tau_2\nonumber\\
&&+v_a\hat{u}^a+v_au^a(\ln\xi\dot{)})
+2(\rho_2+p_2\dot{)}u_av^au_bv^b+4(\rho_2+p_2)(\dot{v}_au^av_bu^b
+v_au^av_bu^b(\ln\xi\dot{)}+v_a\dot{u}^av_bu^b)\nonumber\\
&&=2\beta(\rho_1-p_2+\Lambda+(\rho_2+p_2)v_au^av_bu^b),\\\label{5.6}
&&(\rho_2+p_2\dot{)}(v_av_b+2v_cu^cu_{(a}v_{b)}+v_cu^cv_du^du_au_b+\frac{1}{3}
h_{1ab}-\frac{1}{3}
h_{1ab}v_cu^cv_du^d)+(\rho_1+\rho_2-p_1\nonumber\\
&&-p_2+2{\Lambda})\sigma_{1ab}
+2(\rho_2+p_2)[\dot{v}_{(a}v_{b)}+\dot{v}_cu^cv_{(a}u_{b)}
+\dot{v}_cu^cv_du^du_au_b+\dot{v}_{(a}u_{b)}v_cu^c-\frac{1}{3}h_{1ab}\dot{v}_cu^cv_du^d\nonumber\\
&&+v_cv_{(a}u_{;b)}^c+v_cu^cv_du^d_{;(a}u_{b)}+v_c\dot{u}^cv_{(a}u_{b)}
+v_c\dot{u}^cv_du^du_au_b-\frac{1}{3}h_{1ab}(v_c\hat{u}^c
+v_c\dot{u}^cv_du^d)\nonumber\\
&&+u^cv_cv_{(a}(\ln\xi)_{,b)}+u^cv_c(\ln\xi\dot{)}v_{(a}u_{b)}
+v_cu^cv_du^du_{(a}(\ln\xi)_{,b)}+v_cu^cv_du^d(\ln\xi\dot{)}u_au_b\nonumber
\\&&-\frac{1}{3}h_{1ab}(v_cu^c(\ln\xi\hat{)}+v_cu^cv_du^d(\ln\xi\dot{)})]
=\beta(\rho_2+p_2)(v_a
v_b+2v_cu^cu_{(a}v_{b)}+v_cu^cv_du^du_au_b\nonumber\\
&&+\frac{1}{3} h_{1ab}-\frac{1}{3} h_{1ab}v_cu^cv_du^d).
\end{eqnarray}
If we take $\alpha=0$ in the above relations we find the results for
a RC. Using $1+3$ decomposition of the tensor $P_{ab}$, given in
Eqs.~(\ref{4.6})--(\ref{4.9}), the above system of equations can be
written as follows:
\begin{eqnarray}\label{5.7}
&&\dot{\rho}_1=\dot{p}_2+(\rho_2+p_2)(\hat{v}^a u_a+u_av^a\tau_2)
-(\rho_1+p_1)\theta_1+(\rho_2+p_2\hat{)}u_av^a,\\
&&\dot{p}_1=\frac{1}{3}\dot{\rho}_2-\frac{2}{3}\dot{p}_2
+(\rho_2+p_2)(\hat{v}^au_a+v_au^a\tau_2)+(\rho_2+p_2\hat{)}u^av_a+\frac{1}{3}(2\rho_2
-\rho_1-5p_1-2p_2\nonumber\\&&+4\Lambda)\theta_1+\frac{2}{3}(\rho_2
+p_2\dot{)}u^av_au^bv_b+\frac{4}{3}(\rho_2+p_2)\label{5.8}
\times(\dot{v}_au^av_bu^b+u^av_a(\ln\xi\hat{)}+u^av_au^bv_b(\frac{1}{3}\theta_1
\nonumber\\
&&-\frac{4}{3\xi}\mu_{1P})+v_a\hat{u}^a+v_a\dot{u}^av_bu^b)
-\frac{\beta}{3}(3\rho_1+\rho_2-3p_1-5p_2+6\Lambda+2(\rho_2+p_2)
u_av^au_bv^b),\\\label{5.9}
&&\frac{1}{\xi}(\rho_1-\rho_2+3p_1+p_2-2\Lambda)\Upsilon_{1Pa}
+2(\rho_2+p_2\dot{)}\times(u^cv_cv_a+u^cv_cu^dv_du_a)+2(\rho_2+p_2)(\dot{v}_cu^cv_a\nonumber\\
&&+2\dot{v}_cu^cv_du^du_a+u_cv^cu^dv_d(\ln\xi)_{,a}+2u_cv^cu^dv_du_a
\times(\frac{1}{3}\theta_1-\frac{4}{3\xi}\mu_{1P})+u^cv_cv_a(\frac{1}{3}\theta_1
-\frac{4}{3\xi}\mu_{1P})
\nonumber\\&&+v_cu^cv_du^d_{;a}+2v_c\dot{u}^cv_du^du_a)+v_c\dot{u}^cv_a+v_cu^c\dot{v}_a
=\beta(\rho_2+p_2)(v_cu^cv_a+v_cu^cv_du^du_a),\\\label{5.10}
&&\frac{1}{3}(\rho_2-\rho_1-3p_1-p_2+2\Lambda)(\theta_1+\frac{2}{\xi}\mu_{1P})
+(\rho_2+p_2\hat{)}u^cv_c+(\rho_2+p_2)(\hat{v}^au_a
+u_av^a\tau_2+v_a\hat{u}^a\nonumber\\
&&+v_au^a(\frac{1}{3}\theta_1-\frac{4}{3\xi}\mu_{1P}))
+(\rho_2+p_2\dot{)}u_av^au_bv^b+2(\rho_2+p_2)(\dot{v}_au^av_bu^b
+v_au^av_bu^b(\frac{1}{3}\theta_1-\frac{4}{3\xi}\mu_{1P})\nonumber\\
&&+v_a\dot{u}^av_bu^b)=\beta(\rho_1-p_2+\Lambda+(\rho_2+p_2)v_au^av_bu^b),\\\label{5.11}
&&(\rho_2+p_2\dot{)}(v_a
v_b+2v_cu^cu_{(a}v_{b)}+v_cu^cv_du^du_au_b+\frac{1}{3}h_{1ab}
-\frac{1}{3}h_{1ab}v_cu^cv_du^d)+(\rho_1+\rho_2-p_1-p_2\nonumber\\
&&+2{\Lambda})M_{1P_{ab}}+2(\rho_2+p_2)[\dot{v}_{(a}v_{b)}
+\dot{v}_cu^cv_{(a}u_{b)}
+\dot{v}_cu^cv_du^du_au_b+\dot{v}_{(a}u_{b)}v_cu^c
-\frac{1}{3}h_{1ab}\dot{v}_cu^cv_du^d
\nonumber\\
&&+v_cv_{(a}u_{;b)}^c+v_cu^cv_du^d_{;(a}u_{b)}+v_c\dot{u}^cv_{(a}u_{b)}
+v_c\dot{u}^cv_du^du_au_b-\frac{1}{3}h_{1ab}(v_c\hat{u}^c
+v_c\dot{u}^cv_du^d)+u^cv_cv_{(a}(\ln\xi)_{,b)}\nonumber\\
&&+u^cv_c(\frac{1}{3}\theta_1-\frac{4}{3\xi}\mu_{1P})v_{(a}u_{b)}
+v_cu^cv_du^du_{(a}(\ln\xi)_{,b)}+v_cu^cv_du^d(\frac{1}{3}\theta_1-\frac{4}{3\xi}\mu_{1P})u_au_b-
\frac{1}{3}h_{1ab}(v_cu^c(\ln\xi\hat{)}\nonumber\\
&&+v_cu^cv_du^d(\frac{1}{3}\theta_1-\frac{4}{3\xi}\mu_{1P}))]=\beta(\rho_2+p_2)(v_a
v_b+2v_cu^cu_{(a}v_{b)}+v_cu^cv_du^du_au_b + \frac{1}{3}
h_{1ab}\nonumber\\
&&-\frac{1}{3}h_{1ab}v_cu^cv_du^d).
\end{eqnarray}
These are the necessary and sufficient conditions for a CRC in the
form
of kinematical quantities.\\
\textbf{Proposition 3:} A two perfect fluid spacetime admits a CRC
${\xi}^a={\xi}v^a$ if and only if
\begin{eqnarray}\label{5.12}
&&\dot{\rho}_2=\dot{p}_1+(\rho_1+p_1)(\hat{u}^a v_a+u_av^a\tau_1)
-(\rho_2+p_2)\theta_2+(\rho_1+p_1\hat{)}u_av^a,\\\label{5.13}
&&\dot{p}_2=\frac{1}{3}\dot{\rho}_1-\frac{2}{3}\dot{p}_1
+(\rho_1+p_1)\times(v^a\hat{u}_a+v_au^a\tau_1)+(\rho_1+p_1\hat{)}u^av_a+\frac{1}{3}(2\rho_1-\rho_2\nonumber\\
&&-2p_1
-5p_2+4\Lambda)\theta_2+\frac{2}{3}(\rho_1+p_1\dot{)}u^av_au^bv_b
+\frac{4}{3}(\rho_1+p_1)(\dot{u}_av^av_bu^b+u^av_a(\ln\xi\hat{)}\nonumber\\
&&+u^av_au^bv_b(\ln\xi\dot{)}+u_a\hat{v}^a+u_a\dot{v}^av_bu^b)
-\frac{\beta}{3}(\rho_1+3\rho_2-5p_1-3p_2+6\Lambda+2(\rho_2+p_2)v_au^av_bu^b),\\\label{5.14}
&&(\rho_2-\rho_1+p_1+3p_2-2\Lambda) [\dot{v}_a-(\ln\xi)_{,a}
-(\ln\xi\dot{)}v_a]+2(\rho_1+p_1\dot{)}(u^cv_cu_a+u^cv_cu^dv_dv_a)
\nonumber\\
&&+2(\rho_1+p_1)(\dot{u}_cv^cu_a+2\dot{u}_cv^cv_du^dv_a
+v_cu^c\dot{u}_a+u_cv^cu^dv_d(\ln\xi)_{,a}
+2u_cv^cu^dv_d(\ln\xi\dot{)}v_a
\nonumber\\&&+u^cv_c(\ln\xi\dot{)}u_a+v_cu^cu_dv^d_{;a}
+2u_c\dot{v}^cv_du^dv_a+u_c\dot{v}^cu_a)=\beta(\rho_1+p_1)(v_cu^cu_a+v_cu^cv_du^dv_a),\\\label{5.15}
&&(\rho_1-\rho_2-p_1-3p_2+2\Lambda)[\theta_2-(\ln\xi\dot{)}]+2(\rho_1+p_1\hat{)}u^cv_c+2(\rho_1+p_1)(\hat{u}^av_a
+u_av^a\tau_1\nonumber\\&&+u_a\hat{v}^a+v_au^a(\ln\xi\dot{)})
+2(\rho_1+p_1\dot{)}u_av^au_bv^b+4(\rho_1+p_1)(\dot{u}_av^av_bu^b
+v_au^av_bu^b(\ln\xi\dot{)}\nonumber\\&&+u_a\dot{v}^av_bu^b)=2\beta
(\rho_2-p_1+\Lambda+(\rho_1+p_1)v_au^av_bu^b),\label{5.16}
\end{eqnarray}
\begin{eqnarray}
&&(\rho_1+p_1\dot{)}(u_a
u_b+2v_cu^cu_{(a}v_{b)}+v_cu^cv_du^dv_av_b
+\frac{1}{3}h_{2ab}-\frac{1}{3}h_{2ab}v_cu^cv_du^d)+(\rho_1+\rho_2-
p_1-p_2\nonumber\\
&&+2{\Lambda})\sigma_{2ab}+2(\rho_1+p_1)[\dot{u}_{(a}u_{b)}+\dot{u}_cv^cv_{(a}u_{b)}
+\dot{u}_cv^cv_du^dv_av_b+\dot{u}_{(a}v_{b)}v_cu^c
-\frac{1}{3}h_{2ab}\dot{u}_cv^cv_du^d
\nonumber\\
&&+u_cu_{(a}v_{;b)}^c+v_cu^cu_dv^d_{;(a}v_{b)}+u_c\dot{v}^cv_{(a}u_{b)}
+u_c\dot{v}^cv_du^dv_av_b-\frac{1}{3}h_{2ab}(u_c\hat{v}^c
+u_c\dot{v}^cv_du^d)\nonumber\\
&&+u^cv_cu_{(a}(\ln\xi)_{,b)}+u^cv_c(\ln\xi\dot{)}v_{(a}u_{b)}
+v_cu^cv_du^dv_{(a}(\ln\xi)_{,b)}+v_cu^cv_du^d(\ln\xi\dot{)}v_av_b\nonumber\\
&&-
\frac{1}{3}h_{2ab}(v_cu^c(\ln\xi\hat{)}+v_cu^cv_du^d(\ln\xi\dot{)})]
=\beta(\rho_1+p_1)(u_a u_b+2v_cu^cu_{(a}v_{b)}+v_cu^cv_du^dv_av_b
\nonumber\\&&+ \frac{1}{3}
h_{2ab}-\frac{1}{3}h_{2ab}v_cu^cv_du^d).
\end{eqnarray}
Using the kinematical conditions for the timelike case, i.e.,
Eqs.~(\ref{4.12})--(\ref{4.15}) in Eqs.~(\ref{5.12})--(\ref{5.16}),
we can obtain an expression of the above equations in terms of
$P_{ab}$.

\subsection{Spacelike Conformal Ricci Collineations}

In this case we use the $1+1+2$ decomposition of the Ricci tensor
given in Eq.~(\ref{3.5}) and the corresponding $1+1+2$ expression of
$\pounds_\xi R_{ab}$ given in Eq.~(\ref{3.8}). Equating these two
expressions we find the following results corresponding to the
following two cases:\\
(i)\quad ${\xi}^a={\xi}x^a$,\quad (ii)\quad ${\xi}^a={\xi}y^a$.\\
\textbf{Proposition 4:} A two perfect fluid spacetime admits CRC
$\xi^a=\xi x^a$ if and only if
\begin{eqnarray}\label{5.22}
&&2\rho^*_1-p^*_2+\rho^*_2 +(\rho_1-\rho_2
+3p_1+p_2+2\Lambda)\dot{u}^a x_a+3(\rho_1+\rho_2
-p_1-p_2+2\Lambda)(ln\xi)^*\nonumber\\&&+(\rho_2+p_2)^*u_av^au_bv^b
+2(\rho_2+p_2)(v^*_au^av_bu^b-x_a\dot{v}^av_bu^b)=\beta(2\rho_1
+\rho_2-p_2+2\Lambda)\nonumber\\&&+(\rho_2+p_2)u_av^au_bv^b),\\\label{5.23}
&&(\rho_1+\rho_2-p_1-p_2+2\Lambda)[x^*_a+(\ln\xi)_{,a}-(ln\xi)^*x_a]=0,\\\label{5.24}
&&(\rho_1+\rho_2-p_1-p_2+2\Lambda)(2\dot{u}^a
x_a+\varepsilon_1)+p^*_2+\rho^*_2+2(\rho_2+p_2)\hat{v}^ax_a-(\rho_2+p_2)^*u_av^au_bv^b\nonumber\\
&&-2(\rho_2+p_2)^*v^*_au^av_bu^b+2(\rho_2+p_2)x_a\dot{v}^au^bv_b
=\beta(\rho_2-4p_1-3p_2+4\Lambda-(\rho_2+p_2)u_av^au_bv^b),\\\label{5.25}
&&(\rho_2+p_2)(v^*_cu^cv_a+2v^*_bu^bv_cu^cu_a+v^*_au^cv_c)-(\rho_2+p_2)(x_c
\dot{v}^cv_a+2x_b\dot{v}^bv_cu^cu_a+v_bu^bx_cv^c_{;a})\nonumber\\&&+2(\rho_1-\rho_2+3p_1
+p_2-2\Lambda)\omega_{1af}x^f+(\rho_1+\rho_2-p_1-p_2 +2\Lambda)
N_{1a}+(\rho_2+p_2)^*(u^cv_cv_a\nonumber\\
&&+u^cv_cu^dv_du_a)=\beta(\rho_2+p_2)(v_cu^cv_a
+v_bu^bv_cu^cu_a),\\\label{5.26}
&&(p_2+\rho_2)^*+2(\rho_1+\rho_2-p_1-p_2+2\Lambda) (ln \xi)^*+
2(\rho_2+p_2)x_a\hat{v}^a-(\rho_1+\rho_2-p_1-p_2+2\Lambda)\varepsilon_1
\nonumber\\&&-(\rho_2+p_2)^*v_au^av_bu^b-2(\rho_2+p_2)v^*_au^av_bu^b
+2(\rho_2+p_2)v_au^a\dot{v}^bx_b=\beta(p_2+\rho_2-(\rho_2+p_2)\nonumber\\
&&v_au^av_bu^b),\\\label{5.27}
&&(\rho_2+p_2)^*(v_av_b+2v_cu^cv_{(a}u_{b)}+v_cu^cv_du^du_au_b
+\frac{1}{2}H_{1ab}-\frac{1}{2}H_{1ab}v_cu^cv_du^d)+2(\rho_1+\rho_2-p_1\nonumber\\
&&-p_2+2\Lambda)S_{1ab}+2(\rho_2
+p_2)(v^*_{(a}v_{b)}+v^*_cu^cv_{(a}u_{b)}+v_cu^cv^*_{(a}u_{b)}+v^*_cu^cv_du^du_au_b
-H_{1ab}v^*_cu^cv_du^d\nonumber\\
&&-x_cv_{(a}v_{;b)}^c-v_cu^cx_du_{(a}v^d_{;b)}-x_c\dot{v}^cv_{(a}u_{b)}
-x_c\dot{v}^cv_du^du_au_b+\frac{1}{2}H_{1ab}(x_c\hat{v}^c+x_c\dot{v}^cv_du^d)\nonumber\\
&&=\beta(\rho_2+p_2)(v_av_b+2v_cu^cv_{(a}u_{b)}
+v_cu^cv_du^du_au_b+\frac{1}{2}H_{1ab}-\frac{1}{2}H_{1ab}v_cu^cv_du^d),\\\label{5.28}
&&\dot{\rho_1}+(\rho_1+p_1)\theta_1
-(\rho_2+p_2)(v^a\hat{u}_a+v_au^a\tau_2)-(\rho_2+p_2\hat{)}v_au^a-\dot{p}_2=0,\\\label{5.29}
&&p^*_1+p^*_2+(\rho_1+p_1)\dot{u}_cx^c+
(\rho_2+p_2)\hat{v}_cx^c=0,\\\label{5.30}
&&H^c_{1a}[(\rho_1+p_1)\dot{u}_c+
(\hat{\rho}_2+\hat{p}_2+(\rho_2+p_2)\tau_2)v_c+(\rho_2+p_2)
\hat{v}_c+h^b_{1c}(p_1+p_2)_{;b}]=0.
\end{eqnarray}
We can use the $1+1+2$ decomposition of the tensor $P_{ab}$ given in
Eqs.~(\ref{4.21})--(\ref{4.28}) to write these equations in terms of
the irreducible parts of $P_{ab}$. \\
\textbf{Proposition 5:} A two perfect fluid spacetime admits a CRC
$\xi^a=\xi y^a$ if and only if
\begin{eqnarray}\label{5.40}
&&2\rho^*_2-p^*_1+\rho^*_1 +(\rho_2-\rho_1
+p_1+3p_2+2\Lambda)\dot{v}^c y_c+3(\rho_1+\rho_2
-p_1-p_2+2\Lambda)(ln\xi)^*\nonumber\\
&&+(\rho_1+p_1)^*u_av^au_bv^b+2(\rho_1+p_1)(u^*_av^av_bu^b
-y_a\dot{u}^av_bu^b)=\beta(\rho_1 +2\rho_2-p_1+2\Lambda\nonumber\\
&&+(\rho_1+p_1)u_av^au_bv^b),\\\label{5.41}
&&(\rho_1+\rho_2-p_1-p_2+2\Lambda)[y^*_a+(\ln\xi)_{,a}
-(ln\xi)^*y_a]=0,\\\label{5.42} &&(\rho_1+\rho_2-p_1-p_2
+2\Lambda)(2\dot{v}^a
y_a+\varepsilon_2)+p^*_1+\rho^*_1+2(\rho_1+p_1)\hat{u}^a
y_a-(\rho_2+p_2)^*u_av^au_bv^b\nonumber\\
&&-2(\rho_1+p_1)^*u^*_av^av_bu^b+2(\rho_1+p_1)y_a\dot{u}^au^bv_b=\beta(\rho_1-4p_2-3p_1+4\Lambda-(\rho_1+p_1)u_av^au_bv^b),\\\label{5.43}
&&(\rho_1+p_1)(u^*_cv^cu_a+2u^*_bv^bv_cu^cv_a+u^*_av^cu_c)-(\rho_1+p_1)(y_c
\dot{u}^cu_a+2y_b\dot{u}^bv_cu^cv_a+v_bu^by_cu^c_{;a})\nonumber\\&&+2(\rho_2-\rho_1+p_1
+3p_2-2\Lambda)\omega_{2af}y^f+(\rho_1+\rho_2-p_1-p_2
+2\Lambda) N_{2a}+(\rho_1+p_1)^*(u^cv_cu_a\nonumber\\
&&+u^cv_cu^dv_dv_a)=\beta(\rho_1+p_1)(v_cu^cu_a+v_bu^bv_cu^cv_a),\\\label{5.44}
&&(p_1+\rho_1)^*+2(\rho_1+\rho_2-p_1-p_2+2\Lambda) (ln \xi)^*+
2(\rho_1+p_1)y_a\hat{u}^a-(\rho_1+\rho_2-p_1-p_2+2\Lambda)\varepsilon_2
\nonumber\\&&-(\rho_1+p_1)^*v_au^av_bu^b-2(\rho_1+p_1)u^*_av^av_bu^b
+2(\rho_1+p_1)v_au^a\dot{u}^by_b=\beta(p_1+\rho_1-\nonumber\\
&&(\rho_1+p_1)v_au^av_bu^b),\\\label{5.45}
&&(\rho_1+p_1)^*(u_au_b+2v_cu^cv_{(a}u_{b)}
+v_cu^cv_du^dv_av_b+\frac{1}{2}H_{2ab}
-\frac{1}{2}H_{2ab}v_cu^cv_du^d)+2(\rho_1+\rho_2-p_1\nonumber\\
&&-p_2+2\Lambda)S_{2ab}+2(\rho_1
+p_1)(u^*_{(a}u_{b)}+u^*_cv^cv_{(a}u_{b)}+v_cu^cu^*_{(a}v_{b)}+u^*_cv^cv_du^dv_av_b
-H_{2ab}u^*_cv^cv_du^d\nonumber\\
&&-y_c\dot{u}^cv_du^dv_av_b-y_cu_{(a}u_{;b)}^c
-v_cu^cy_dv_{(a}u^d_{;b)}-y_c\dot{u}^cv_{(a}u_{b)}+
\frac{1}{2}H_{2ab}(y_c\hat{u}^c+y_c\dot{u}^cv_du^d)=\beta(\rho_1+p_1)\nonumber\\
&&(u_au_b+2v_cu^cv_{(a}u_{b)}
+v_cu^cv_du^dv_av_b+\frac{1}{2}H_{2ab}-\frac{1}{2}H_{2ab}v_cu^cv_du^d),\\\label{5.46}
&&\dot{\rho_2}+(\rho_2+p_2)\theta_2
-(\rho_1+p_1)(u^a\hat{v}_a+v_au^a\tau_1)-(\rho_1+p_1\hat{)}v_au^a-\dot{p}_1=0,\\\label{5.47}
&&p^*_1+p^*_2+(\rho_2+p_2)\dot{v}_cy^c+
(\rho_1+p_1)\hat{u}_cy^c=0,\\\label{5.48}
&&H^c_{2a}[(\rho_2+p_2)\dot{v}_c+
(\hat{\rho}_1+\hat{p}_1+(\rho_1+p_1)\tau_1)u_c+(\rho_1+p_1)\hat{u}_c+h^b_{2c}(p_1+p_2)_{;b}]=0.
\end{eqnarray}
We can use the $1+1+2$ decomposition of the tensor $P_{ab}$ given in
Eqs.~(\ref{4.31})--(\ref{4.38}) to write these equations in terms of
the irreducible parts of $P_{ab}$.

\section{Conformal Matter Collineations}

\subsection{Timelike Conformal Matter Collineations}

Here we give the necessary and sufficient conditions for the
timelike CMCs of the two perfect fluids for the following cases:\\
\par\noindent
(i)\quad ${\xi}^a={\xi}u^a$,\quad (ii)\quad ${\xi}^a={\xi}v^a$.\\
\textbf{Proposition 6:} A two perfect fluid admits a CMC
${\xi}^a={\xi}u^a$ if and only if
\begin{eqnarray}\label{6.1}
&&(p_2-\rho_1)[\dot{u}_a-(ln \xi)_{, a}
-u_a(\ln\xi\dot{)}]+(\rho_2+p_2\dot{)}(u^cv_cu^dv_du_a+u^cv_cv_a)+(\rho_2+p_2)(\dot{v}_cu^cv_a\nonumber\\
&&+2\dot{v}_cu^cu^dv_du_a+u^cv_c\dot{v}_a
+v_c\dot{u}^cv_a+v_cu^cv_du^d_{;a}+2\dot{u}_cv^cu^dv_du_a
+2(\ln\xi\dot{)}v_cu^cu^dv_du_a\nonumber\\
&&+(\ln\xi\dot{)}v_au^cv_c+u_cv^cu_dv^d(\ln\xi)_{,a})
=\beta(\rho_2+p_2)(v_cu^cv_a+v_cu^cv_du^du_a),\\\label{6.2}
&&\dot{p}_1+\frac{2}{3}\dot{p}_2-\frac{1}{3}
\dot{\rho}_2+\frac{2}{3}(p_1+p_2)\theta_1+\frac{1}{3}(\rho_2
+p_2\dot{)}v_au^av_bu^b+\frac{2}{3}(\rho_2+p_2)(v_a\hat{u}^a+\dot{v}_au^av_bu^b\nonumber\\
&&+v_a\dot{u}^av_bu^b +u^av_a(\ln \xi
\hat{)}+u_av^av_bu^b(\ln\xi\dot{)})=\beta(p_1+\frac{2}{3}p_2
-\frac{1}{3}\rho_2+\frac{1}{3}(\rho_2+p_2)u_av^au_bv^b),\\\label{6.3}
 &&2(\rho_1-p_2)(\ln
\xi\dot{)}-(\rho_1+p_1)\theta_1
+(\rho_2+p_2)(u^a\hat{v}_a+u_av^a\tau_1+2\dot{v}_au^av_bu^b+2u^av_au^bv_b(\ln\xi\dot{)}
\nonumber\\
&&+2v_a\dot{u}^av_bu^b)+(\rho_2+p_2\dot{)}u^av_au^bv_b+(\rho_2+p_2\hat{)}v^au_a
=\beta(\rho_1-p_2+(\rho_2+p_2)u^av_au^bv_b),\label{6.4}
\end{eqnarray}
\begin{eqnarray}
&&(\rho_2+p_2\dot{)}(v_a
v_b+2v_cu^cu_{(a}v_{b)}+v_cu^cv_du^du_au_b+\frac{1}{3}h_{1ab}-\frac{1}{3}h_{1ab}v_cu^cv_du^d)+2(p_1+p_2)\sigma_{1ab}\nonumber\\
&&+2(\rho_2+p_2)[\dot{v}_{(a}v_{b)}+\dot{v}_cu^cv_{(a}u_{b)}
+\dot{v}_{(a}u_{b)}v_cu^c+\dot{v}_cu^cv_du^du_au_b
-\frac{1}{3}h_{1ab}\dot{v}_cu^cv_du^d
+v_cv_{(a}u_{;b)}^c\nonumber\\
&&+v_cu^cv_du^d_{;(a}u_{b)}+v_c\dot{u}^cv_{(a}u_{b)}+v_c\dot{u}^cv_du^du_au_b-\frac{1}{3}h_{1ab}(v_c\hat{u}^c
+v_c\dot{u}^cv_du^d)+u^cv_cv_{(a}(\ln\xi)_{,b)}\nonumber\\
&&+u^cv_c(\ln\xi\dot{)}v_{(a}u_{b)}
+v_cu^cv_du^du_{(a}(\ln\xi)_{,b)}+v_cu^cv_du^d(\ln\xi\dot{)}u_au_b
-\frac{1}{3}h_{1ab}(v_cu^c(\ln\xi\hat{)}\nonumber\\
&&+v_cu^cv_du^d(\ln\xi\dot{)})]=\beta(\rho_2+p_2)(v_a v_b+2v_cu^cu_{(a}v_{b)}+v_cu^cv_du^du_au_b\nonumber\\
&&+\frac{1}{3} h_{1ab}-\frac{1}{3}
h_{1ab}v_cu^cv_du^d).
\end{eqnarray}
The {\it energy} conservation equation (\ref{4.9}) remains the same
\begin{eqnarray}\label{6.5}
&&\dot{\rho_1}+(\rho_1+p_1)\theta_1
-(\rho_2+p_2)u^a\hat{v}_a-\dot{p}_2-(\rho_2+p_2\hat{)}v^au_a
-(\rho_2+p_2)u_av^a\tau_1=0.
\end{eqnarray}
If we take $\alpha=0$ in the above relations we get result for the
matter collineation. We can use the $1+3$ decomposition of the
tensor $P_{ab}$ given in Eq.~(\ref{4.6})--(\ref{4.9}) to write these
equations in terms of
irreducible parts of $P_{ab}$.\\
\textbf{Proposition 7:} A two perfect fluid solution admits a CMC
${\xi}^a={\xi}v^a$ if and only if
\begin{eqnarray}\label{6.11}
&&\dot{\rho_2}+(\rho_2+p_2)\theta_2-(\rho_1+p_1)v^a\hat{u}_a
-\dot{p}_1-(\rho_1+p_1\hat{)}v^au_a-(\rho_1+p_1)u_av^a\tau_2=0,\\\label{6.12}
&&(p_1-\rho_2)[\dot{v}_a-(\ln \xi)_{, a} -v_a(\ln\xi\dot{)}]
+(\rho_1+p_1\dot{)}(u^cv_cu^dv_dv_a+u^cv_cu_a)+(\rho_1+p_1)\nonumber\\
&&\times(\dot{u}_cv^cv_a+2\dot{u}_cv^cu^dv_dv_a+u^cv_c\dot{u}_a
+u_c\dot{v}^cu_a+v_cu^cu_dv^d_{;a}+2\dot{v}_cu^cu^dv_dv_a
+2(\ln\xi\dot{)}\nonumber\\
&&v_cu^cu^dv_dv_a+(\ln\xi\dot{)}v_au^cu_c+u_cv^cu_dv^d(\ln\xi)_{,a})
=\beta(\rho_1+p_1)\times(v_cu^cu_a+v_cu^cv_du^dv_a),\\\label{6.13}
&&\dot{p}_2+\frac{2}{3}\dot{p}_1-\frac{1}{3}
\dot{\rho}_1+\frac{2}{3}(p_1+p_2)\theta_2 +\frac{1}{3}(\rho_1
+p_1\dot{)}v_au^av_bu^b+\frac{2}{3}(\rho_1+p_1)(u_a\hat{v}^a+\dot{u}_av^av_bu^b\nonumber\\
&&+u_a\dot{v}^av_bu^b +u^av_a(\ln \xi
\hat{)}+u_av^av_bu^b(\ln\xi\dot{)})=\beta(p_2+\frac{2}{3}p_1
-\frac{1}{3}\rho_1+\frac{1}{3}(\rho_1+p_1)u_av^au_bv^b),\\\label{6.14}
&&2(\rho_2-p_1)(\ln \xi\dot{)}-(\rho_2+p_2)\theta_2
+(\rho_1+p_1)\times(v^a\hat{u}_a+u_av^a\tau_2+2\dot{u}_av^av_bu^b+2u^av_au^bv_b(\ln\xi\dot{)}
\nonumber\\
&&+2u_a\dot{v}^av_bu^b)+(\rho_1+p_1\dot{)}u^av_au^bv_b+(\rho_1+p_1\hat{)}v^au_a
=\beta(\rho_2-p_1+(\rho_1+p_1)u^av_au^bv_b),\\\label{6.15}
&&(\rho_1+p_1\dot{)}(u_a
u_b+2v_cu^cu_{(a}v_{b)}+v_cu^cv_du^dv_av_b +\frac{1}{3}h_{2ab}-\frac{1}{3}h_{2ab}v_cu^cv_du^d)+2(p_1+p_2)\sigma_{2ab}\nonumber\\
&&+2(\rho_1+p_1)[\dot{u}_{(a}u_{b)}+\dot{u}_cv^cv_{(a}u_{b)}
+\dot{u}_{(a}v_{b)}v_cu^c+\dot{u}_cv^cv_du^dv_av_b
-\frac{1}{3}h_{2ab}\dot{u}_cv^cv_du^d+u_cu_{(a}v_{;b)}^c\nonumber\\
&&+v_cu^cu_dv^d_{;(a}v_{b)}+u_c\dot{v}^cv_{(a}u_{b)}+u_c\dot{v}^cv_du^dv_av_b-\frac{1}{3}h_{2ab}(u_c\hat{v}^c
+u_c\dot{v}^cv_du^d)+u^cv_cu_{(a}(\ln\xi)_{,b)}\nonumber\\
&&+u^cv_c(\ln\xi\dot{)}u_{(a}v_{b)}+v_cu^cv_du^dv_{(a}(\ln\xi)_{,b)}+v_cu^cv_du^d(\ln\xi\dot{)}v_av_b
-\frac{1}{3}h_{2ab}(v_cu^c(\ln\xi\hat{)}+v_cu^cv_du^d(\ln\xi\dot{)})]\nonumber\\
&&=\beta(\rho_1+p_1)(u_a u_b+2v_cu^cu_{(a}v_{b)}+\frac{1}{3}
h_{2ab}+v_cu^cv_du^dv_av_b-\frac{1}{3} h_{2ab}v_cu^cv_du^d).
\end{eqnarray}
We can write these equations by using the $1+3$ decomposition of the
tensor $P_{ab}$ given in Eqs.~(\ref{4.12})--(\ref{4.15}).

\subsection{Spacelike Conformal Matter Collineations}

The necessary and sufficient conditions for the timelike CMCs of
the two perfect fluids are given for the following two cases:\\
\par\noindent
(i)\quad ${\xi}^a={\xi}x^a$,\quad (ii)\quad ${\xi}^a={\xi}y^a$.\\
\textbf{Proposition 8:} A two perfect fluid spacetime admits the CMC
$\xi^a=\xi x^a$ iff
\begin{eqnarray}\label{6.21}
&&(\rho_1-p_2)^*+(\rho_2+p_2)^*v_au^av_bu^b+2(\rho_2+p_2)v^*_au^av_bu^b+2(\rho_1-p_2)\dot{u}^ax_a-(\rho_2+p_2)
\dot{v}^ax_av_bu^b\nonumber\\&&=\beta[(\rho_1-p_2)+(\rho_2+p_2)v_au^av_bu^b],\label{6.22}
\end{eqnarray}
\begin{eqnarray}
&&(p_1+p_2)[x^*_a+(\ln\xi)_{,a}-(ln \xi)^* x_a)=0,\\\label{6.23}
&&(\rho_2+p_2)^*-2(p_1+p_2)\varepsilon_1+2(\rho_1+p_1)\dot{u}^ax_a
+4(\rho_2+p_2)x_c\hat{v}^c-(\rho_2+p_2)^*v_au^av_bu^b\nonumber\\
&&-2(\rho_2+p_2)v^*_au^av_bu^b
+2(\rho_2+p_2)v_au^a\dot{v}^bx_b=\beta(\rho_2-p_2-2p_1-(\rho_2+p_2)v_au^av_bu^b),\\\label{6.24}
&&(\rho_2+p_2)(v^* _c u^c v_a +2v^*_cu^cv_du^du_a+u_cv^cv^*_a
+x_cv^c_{;a}v_du^d+2x_c\dot{v}^cv_du^du_a+x_c\dot{v}^cv_a)\nonumber\\&&+2(\rho_1-p_2)\omega_{1af}
x^f+(p_1+p_2)N_{1a}
=\beta(\rho_2+p_2)(u^cv_cv_a+u^cv_cu^dv_du_a),\\\label{6.25}
&&(p_2+\rho_2)^*+4(p_1+p_2)(\ln \xi)^*+2(\rho_2+p_2)x_a\hat{v}^a-(\rho_2+p_2)^*v_au^av_bu^b+2(p_1+p_2)\varepsilon_1\nonumber\\
&&-2(\rho_2+p_2)v^*_au^av_bu^b
+2(\rho_2+p_2)v_au^a\dot{v}^bx_b=\beta(p_2+\rho_2-(\rho_2+p_2)v_au^av_bu^b),\\\label{6.26}
&&(\rho_2+p_2)^*(v_av_b+2v_cu^cv_{(a}u_{b)}
+v_cu^cv_du^du_au_b+\frac{1}{2}H_{1ab}
-\frac{1}{2}H_{1ab}v_cu^cv_du^d)+2(p_1+p_2)S_{1ab}\nonumber\\
&&+2(\rho_2
+p_2)(v^*_{(a}v_{b)}+v^*_cu^cv_{(a}u_{b)}+v^*_cu^cv_du^du_au_b+v_cu^cv^*_{(a}u_{b)}
-H_{1ab}v^*_cu^cv_du^d\-x_c\dot{v}^cv_{(a}u_{b)}\nonumber\\&&
-x_c\dot{v}^cv_du^du_au_b-x_cv_{(a}v_{;b)}^c
-v_cu^cx_du_{(a}v^d_{;b)}+\frac{1}{2}H_{1ab}(x_c\hat{v}^c+x_c\dot{v}^cv_du^d)
=\beta(\rho_2+p_2)(v_av_b\nonumber\\&&+2v_cu^cv_{(a}u_{b)}
+v_cu^cv_du^du_au_b+\frac{1}{2}H_{1ab}-\frac{1}{2}H_{1ab}v_cu^cv_du^d),\\\label{6.27}
&&\dot{\rho_1}+(\rho_1+p_1)\theta_1
-(\rho_2+p_2)(v^a\hat{u}_a+v_au^a\tau_2)-(\rho_2+p_2\hat{)}v_au^a-\dot{p}_2=0,\\\label{6.28}
&&p^*_1+p^*_2+(\rho_1+p_1)\dot{u}_cx^c+
(\rho_2+p_2)\hat{v}_cx^c=0,\\\label{6.29}
&&H^c_{1a}[(\rho_1+p_1)\dot{u}_c+
(\hat{\rho}_2+\hat{p}_2+(\rho_2+p_2)\tau_2)v_c+(\rho_2+p_2)
\hat{v}_c+h^b_{1c}(p_1+p_2)_{;b}]=0.
\end{eqnarray}
These equations can be written in terms of the irreducible parts of
$P_{ab}$ using the $1+1+2$ decomposition of the tensor $P_{ab}$.\\
\textbf{Proposition 9:} A two perfect fluid spacetime admits the CMC
$\xi^a=\xi y^a$ iff
\begin{eqnarray}\label{6.39}
&&(\rho_2-p_1)^*+(\rho_1+p_1)^*v_au^av_bu^b+2(\rho_1+p_1)u^*_av^av_bu^b+2(\rho_2-p_1)\dot{v}^ay_a
-(\rho_1+p_1)\dot{u}^ay_av_bu^b\nonumber\\&&=\beta[(\rho_2-p_1)+(\rho_1+p_1)v_au^av_bu^b],\\\label{6.40}
&&(p_1+p_2)[y^*_a+(\ln\xi)_{,a}-(ln \xi)^* y_a)=0,\\\label{6.41}
&&(\rho_1+p_1)^*-2(p_1+p_2)\varepsilon_2+2(\rho_2+p_2)\dot{v}^ay_a
+4(\rho_1+p_1)y_a\hat{u}^a-(\rho_1+p_1)^*v_au^av_bu^b\nonumber\\
&&-2(\rho_1+p_1)u^*_av^av_bu^b
+2(\rho_1+p_1)v_au^a\dot{u}^by_b=\beta(\rho_1-p_1-2p_2-(\rho_1+p_1)v_au^av_bu^b),\\\label{6.42}
&&(\rho_1+p_1)(u^* _c v^c u_a +2u^*_cv^cv_du^dv_a+u_cv^cu^*_a
+y_cu^c_{;a}v_du^d+2y_c\dot{u}^cv_du^dv_a+y_c\dot{u}^cu_a)\nonumber\\
&&+2(\rho_2-p_1)\omega_{2af}
y^f+(p_1+p_2)N_{2a}=\beta(\rho_1+p_1)(u^cv_cu_a+u^cv_cu^dv_dv_a),\\\label{6.43}
&&(p_1+\rho_1)^*+4(p_1+p_2)(\ln
\xi)^*+2(\rho_1+p_1)y_a\hat{u}^a-(\rho_1+p_1)^*v_au^av_bu^b
-2(p_1+p_2)\varepsilon_2\nonumber\\&&-2(\rho_1+p_1)u^*_av^av_bu^b+2(\rho_1+p_1)v_au^a\dot{u}^by_b=\beta(p_1+\rho_1
-(\rho_1+p_1)v_au^av_bu^b),\\\label{6.44}
&&(\rho_1+p_1)^*(u_au_b+2v_cu^cv_{(a}u_{b)}+v_cu^cv_du^dv_av_b
+\frac{1}{2}H_{2ab}-\frac{1}{2}H_{2ab}v_cu^cv_du^d)+2(p_1+p_2)S_{2ab}\nonumber\\
&&+2(\rho_1+p_1)(u^*_{(a}u_{b)}+u^*_cv^cv_{(a}u_{b)}+u^*_cv^cv_du^dv_av_b+v_cu^cu^*_{(a}v_{b)}
-H_{2ab}u^*_cv^cv_du^d-v_cu^cy_dv_{(a}u^d_{;b)}\nonumber\\
&&-y_c\dot{u}^cv_{(a}u_{b)}-y_cu_{(a}u_{;b)}^c
-y_c\dot{u}^cv_du^dv_av_b+\frac{1}{2}H_{2ab}(y_c\hat{u}^c+y_c\dot{u}^cv_du^d)=\beta
(\rho_1+p_1)(u_au_b+2v_cu^cv_{(a}u_{b)}\nonumber\\&&+v_cu^cv_du^dv_av_b+\frac{1}{2}H_{2ab}
-\frac{1}{2}H_{2ab}v_cu^cv_du^d),\\\label{6.45}
&&\dot{\rho_2}+(\rho_2+p_2)\theta_2
-(\rho_1+p_1)(u^a\hat{v}_a+v_au^a\tau_1)-(\rho_1+p_1\hat{)}v_au^a-\dot{p}_1=0,\\\label{6.46}
&&p^*_1+p^*_2+(\rho_2+p_2)\dot{v}_cy^c+
(\rho_1+p_1)\hat{u}_cy^c=0,\\\label{6.47}
&&H^c_{2a}[(\rho_2+p_2)\dot{v}_c+
(\hat{\rho}_1+\hat{p}_1+(\rho_1+p_1)\tau_1)u_c+(\rho_1+p_1)
\hat{u}_c+h^b_{2c}(p_1+p_2)_{;b}]=0.
\end{eqnarray}
We can use the $1+1+2$ decomposition of the tensor $P_{ab}$ given in
Eqs.~(\ref{4.31})--(\ref{4.38}) to write these equations in terms of
the irreducible parts of $P_{ab}$.

\section{Outlook}

This section contains a summary and discussion of the results
obtained. We have studied two types of collineations for the
combination of the two perfect fluids. We have derived conditions
for the existence of CRCs and CMCs. In terms of the kinematic
quantities of the vector fields, these conditions are used to
calculate kinematical effects. Using the kinematic and the dynamic
equations, we have obtained the set of equations under the symmetry
assumption. For the physical system, the full set of equations is
the set of these equations plus the Einstein field equations and any
other equations for the extra fields or other geometrical
identities. In the case of timelike CRCs and CMCs, we have found
five conditions for the two perfect fluids. In the case of spacelike
CRCs and CMCs, we have obtained nine conditions. In the following,
we discuss some special cases of the fluid spacetimes.
\begin{description}
\item{1.} When either $\rho_1=0=p_1$ or $\rho_2=0=p_2$, the conditions
of timelike CRC become
\begin{eqnarray}\label{7.1}
&&((\rho_1+3p_1-2\Lambda)\xi
u^a)_{;a}=-\beta(\rho_1\nonumber\\&&-3p_1+4\Lambda),\\\label{7.2}
&&(\rho_1+3p_1-2\Lambda)[\dot{u}_a-(\ln\xi)_{,a}
-\theta_1u_a]\nonumber\\&&=2\beta (\rho_1+\Lambda)u_a,\\\label{7.3}
&&(\rho_1-p_1+2\Lambda)\sigma_{1ab}=0.
\end{eqnarray}
These are the same conditions as for the timelike CRC perfect fluid
\cite{17}.
\item{2.} If $p_1=0=p_2$, this reduces to the case of two
dusts.
\item{3.} If we substitute $\rho_1=p_1$ and $\rho_2=p_2$ in the two fluids,
the respective conditions reduce to the conditions of stiff matter.
\item{4.} For $\rho_1=3p_1,~\rho_2=3p_2$ we have the radiation
case.
\item{5.} When $\rho_1=-p_1,~\rho_2=-p_2$, it
gives a dark energy component.
\end{description} Similarly, we can classify
spacelike CRCs.

If we take either $\rho_1=0=p_1$ or $\rho_2=0=p_2$, the conditions
for the existence of timelike CMCs become
\begin{eqnarray}\label{7.4}
&&\dot{\rho_1}+(\rho_1+p_1)\theta=0,\\\label{7.5}
&&\rho_1(\dot{u}_a-(\ln \xi)_{,a}-u_a(\ln
\xi\dot{)})=0,\\\label{7.6} &&3\dot{p}_1+2p_1\theta_1=3\beta
p_1,\\\label{7.7} &&2p_1\sigma_{1ab}=0,\\\label{7.8} &&2\rho_1(\ln
\xi\dot{)}-p_1\theta=\beta \rho_1.
\end{eqnarray}
These are the conditions for a perfect fluid already available in
the literature \cite{14}. Similarly, we can write down the
conditions for the timelike and spacelike CMCs two dust, stiff
matter, radiation, and dark energy component. It is interesting to
note that if we replace $\frac{1}{2}(\rho_1+3p_1-\Lambda)$ by
$\rho_1$, $\frac{1}{2}(\rho_2+3p_2-\Lambda)$ by $\rho_2$ and
$\frac{1}{2}(\rho_1-p_1+\Lambda)$ by $p_1$,
$\frac{1}{2}(\rho_2-p_2+\Lambda)$ by $p_2$ and vice versa, then we
can obtain the conditions of CMCs from CRCs and vice versa. Also, it
is important to note that the results obtained in the case (i) for
the two perfect fluids (in terms of $u_a$ and $x_a$) can be modified
for the case (ii) by substituting $v_a$ and $y_a$ instead of $u_a$
and $x_a$. It is worth mentioning that when $\alpha=0$, the
conditions for the existence of CRCs and CMCs in all the cases of
the two perfect fluids reduce to the conditions for the existence of
RCs and MCs, respectively \cite{14,17}.

%\end{multicols}
\end{document}